\documentclass[%
reprint,
twocolumn,
notitlepage,
superscriptaddress,
jmp,%
amsmath,amssymb,
 prx,
letter,%
]{revtex4-2}
\usepackage{graphicx}
\usepackage{subfigure}
\usepackage{dcolumn}
\usepackage{bm}
\usepackage[mathlines]{lineno}
\graphicspath{{figure/}}
\usepackage{array}
\usepackage{braket}
\usepackage{diagbox}
\usepackage{amsmath}
\usepackage{appendix}
\usepackage{multirow}
\usepackage{bm,color,bbm}
\usepackage{float}
\newcolumntype{.}{D{.}{.}{-1}}
\usepackage{amsopn}
\usepackage[colorlinks,linkcolor=blue, anchorcolor=blue, urlcolor=blue, citecolor=blue]{hyperref}
\usepackage{epstopdf}
\usepackage{amssymb}
\usepackage{upgreek}
\usepackage{lipsum}

\begin{document}

\title{Field demonstration of distributed quantum sensing without post-selection}

\author{Si-Ran Zhao}
\thanks{These authors contributed equally to this work.}

\author{Yu-Zhe Zhang}
\thanks{These authors contributed equally to this work.}

\author{Wen-Zhao Liu}

\author{Jian-Yu Guan}
\affiliation{Hefei National Laboratory for Physical Sciences at the Microscale and Department of Modern Physics, University of Science and Technology of China, Hefei 230026, China}
\affiliation{Shanghai Branch, CAS Center for Excellence in Quantum Information and Quantum Physics, University of Science and Technology of China, Shanghai 201315, China}
\affiliation{Shanghai Research Center for Quantum Sciences, Shanghai 201315, China}

\author{Weijun Zhang}
\affiliation{State Key Laboratory of Functional Materials for Informatics, Shanghai Institute of Microsystem and Information Technology, Chinese Academy of Sciences, Shanghai 200050, P.~R.~China}

\author{Cheng-Long Li}

\author{Bing Bai}

\author{Ming-Han Li}

\author{Yang Liu}

\affiliation{Hefei National Laboratory for Physical Sciences at the Microscale and Department of Modern Physics, University of Science and Technology of China, Hefei 230026, China}
\affiliation{Shanghai Branch, CAS Center for Excellence in Quantum Information and Quantum Physics, University of Science and Technology of China, Shanghai 201315, China}
\affiliation{Shanghai Research Center for Quantum Sciences, Shanghai 201315, China}

\author{Lixing You}
\affiliation{State Key Laboratory of Functional Materials for Informatics, Shanghai Institute of Microsystem and Information Technology, Chinese Academy of Sciences, Shanghai 200050, P.~R.~China}

\author{Jun Zhang}

\affiliation{Hefei National Laboratory for Physical Sciences at the Microscale and Department of Modern Physics, University of Science and Technology of China, Hefei 230026, China}
\affiliation{Shanghai Branch, CAS Center for Excellence in Quantum Information and Quantum Physics, University of Science and Technology of China, Shanghai 201315, China}
\affiliation{Shanghai Research Center for Quantum Sciences, Shanghai 201315, China}

\author{Jingyun Fan}
\affiliation{Hefei National Laboratory for Physical Sciences at the Microscale and Department of Modern Physics, University of Science and Technology of China, Hefei 230026, China}
\affiliation{Shanghai Branch, CAS Center for Excellence in Quantum Information and Quantum Physics, University of Science and Technology of China, Shanghai 201315, China}
\affiliation{Shanghai Research Center for Quantum Sciences, Shanghai 201315, China}
\affiliation{Shenzhen Institute for Quantum Science and Engineering and Department of Physics, Southern University of Science and Technology, Shenzhen, 518055, P.~R.~China }

\author{Feihu Xu}

\author{Qiang Zhang}

\author{Jian-Wei Pan}

\affiliation{Hefei National Laboratory for Physical Sciences at the Microscale and Department of Modern Physics, University of Science and Technology of China, Hefei 230026, China}
\affiliation{Shanghai Branch, CAS Center for Excellence in Quantum Information and Quantum Physics, University of Science and Technology of China, Shanghai 201315, China}
\affiliation{Shanghai Research Center for Quantum Sciences, Shanghai 201315, China}

\begin{abstract}
Distributed quantum sensing can provide quantum-enhanced sensitivity beyond the shot-noise limit (SNL) for sensing spatially distributed parameters. To date, distributed quantum sensing experiments have been mostly accomplished in laboratory environments without a real space-separation for the sensors. In addition, the post-selection is normally assumed to demonstrate the sensitivity advantage over the SNL. Here, we demonstrate distributed quantum sensing in field and show the \emph{unconditional} violation (without post-selection) of SNL up to $0.916$ dB for the field distance of $240$ m. The achievement is based on a loophole free Bell test setup with entangled photon pairs at the averaged heralding efficiency of $73.88\%$. Moreover, to test quantum sensing in real life, we demonstrate the experiment for long distances (with 10-km fiber) together with the sensing of a completely random and unknown parameter. The results represent an important step towards a practical quantum sensing network for widespread applications.
\end{abstract}

\maketitle

\emph{Introduction.}
By exploiting the quantum mechanical effects, quantum metrology can provide superior sensitivity compared to classical strategies \cite{giovannetti2011advances}. Its sensitivity can surpass the shot-noise limit (SNL) or even reach the Heisenberg limit \cite{kok2002creation,giovannetti2004quantum}, which is the maximum sensitivity bound optimized over all possible quantum states. Considerable efforts have been made to harness different types of quantum resources \cite{degen2017quantum,braun2018quantum} such as the entangled N00N state \cite{walther2004broglie,mitchell2004super,nagata2007beating,resch2007time,dowling2008quantum}, squeezed state \cite{aasi2013enhanced} and quantum coherence \cite{higgins2007entanglement,daryanoosh2018experimental}.

Distributed metrology has attracted considerable attention for applications \cite{brida2010experimental,PhysRevLett.109.123601,PhysRevLett.116.030801,komar2014quantum}. In most of the applications, the distributed sensors are encoded with independent parameters and collectively processed to estimate the linear combination of multiple parameters \cite{PhysRevLett.120.080501,PhysRevLett.121.043604}. Recently, it has been shown by both theories \cite{PhysRevLett.120.080501,PhysRevLett.121.130503,PhysRevLett.121.043604,PhysRevLett.111.070403,PhysRevLett.119.130504} and experiments \cite{guo2020distributed,PhysRevLett.124.150502} that the sensitivity of sensing networks can be considerably improved using entanglement among distributed sensors.

So far, the experiments on quantum metrology have been largely demonstrated in laboratories \cite{walther2004broglie,mitchell2004super,nagata2007beating,resch2007time,higgins2007entanglement,daryanoosh2018experimental}, including the recent distributed quantum metrology demonstrations \cite{guo2020distributed,PhysRevLett.124.150502}. A real-world distributed quantum sensing network in the field has not been implemented yet. Furthermore, the experiments to show the sensitivity advantage were operated under the assumption of post-selection \cite{walther2004broglie,mitchell2004super,nagata2007beating,resch2007time,higgins2007entanglement,daryanoosh2018experimental}.
The only exception is the remarkable work by Slussarenko et al. \cite{slussarenko2017unconditional} which demonstrated the \emph{unconditional} violation of SNL \cite{resch2007time,nagata2007beating} with a single sensor in a laboratory. The unconditional violation means that the imperfections and losses of the system are considered, i.e., all the photons used should be taken into account without post-selection \cite{resch2007time,nagata2007beating}. Nonetheless, a quantum sensing network with multiple distributed  sensors that can unconditionally beat the SNL remains to be solved. Note that the setup of entangled state generation in ref. \cite{slussarenko2017unconditional} can not be used to realize distributed quantum sensing.

In this letter, we report a field test of distributed quantum sensing, which unconditionally beats the SNL. The linear function of two phase parameters is realized with two entangled photons, where one of the photons passes the phase shifts twice. Our experiment demonstrates the state-of-the-art averaged heralding efficiency of $73.88\%$ \cite{christensen2013detection,giustina2015significant,shalm2015strong,li2018test} and achieves a phase precision of $0.916$ dB below the SNL for the field distance of $240$ m.
In addition, we show that the setup can perform distributed quantum sensing using up to $10$-km fiber spools. Moreover, to simulate the actual working circumstances where the phase of each sensor may vary along with the environmental random variations, we use a quantum random number generator to introduce random phase changes and precisely measure the unknown phases.

\begin{figure*}
	\centering
	\includegraphics[width=1\textwidth]{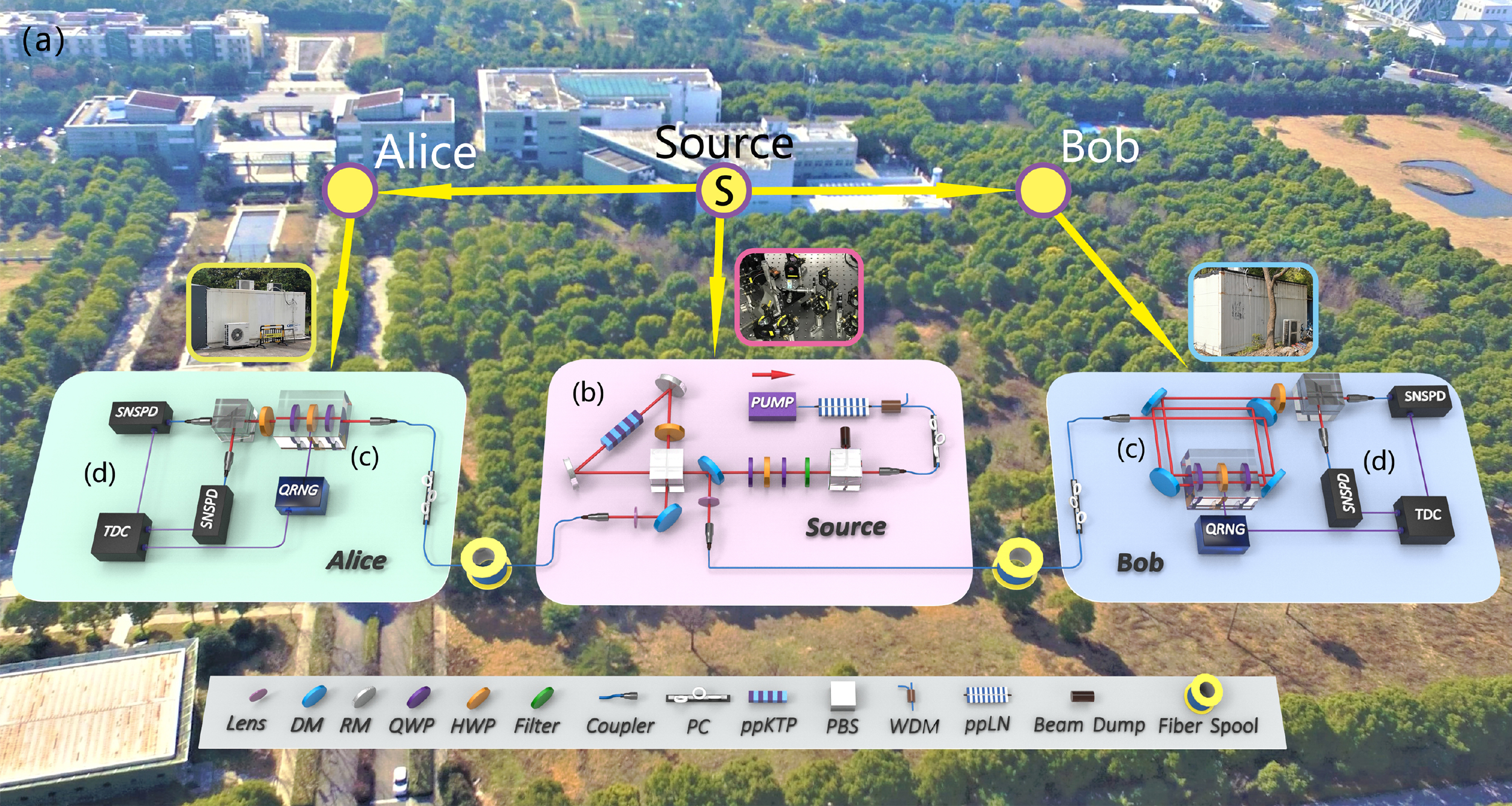}
	\caption{Schematics of the experiment. \textbf{(a)} A bird's-eye view of the satellitic experimental layout. Alice and Bob are on the different sides of the entanglement source, and the linear distance between Alice (Bob) and source is $93 \pm1 (90 \pm1)$ m. \textbf{(b)} Creation of pairs of entangled photons: By injecting a $1560$-nm-wavelength laser into the periodically poled lithium niobate (PPLN) waveguide, the wavelength of generated photons is upconverted to $780$ nm. The pump photons of $780$ nm are injected into the periodically poled potassium titanyl phosphate (PPKTP) crystal in the Sagnac loop to generate polarization-entangled photon pairs of $1560$-nm-wavelength. Then, the photons are sent by fibers in opposite directions to Alice and Bob for different phase shifts and measurements. \textbf{(c)} Realization of different phase shifts: With the combination of QWP, HWP, and QWP, phase shift $\theta$ is implemented between two optical axes. HWP can be controlled by a quantum random number generator (QRNG) (see the main text for more details). \textbf{(d)} Single photon polarization measurement: Photons are projected into one of the $\sigma_x$ bases and are, then, detected by superconducting nanowire single photon detectors (SNSPDs). The time-digital convertor (TDC) is applied to record the single photon detection and random number generation events.}
\label{fig:setup}
\end{figure*}

\begin{figure}
	\centering
	\includegraphics[width=0.45\textwidth]{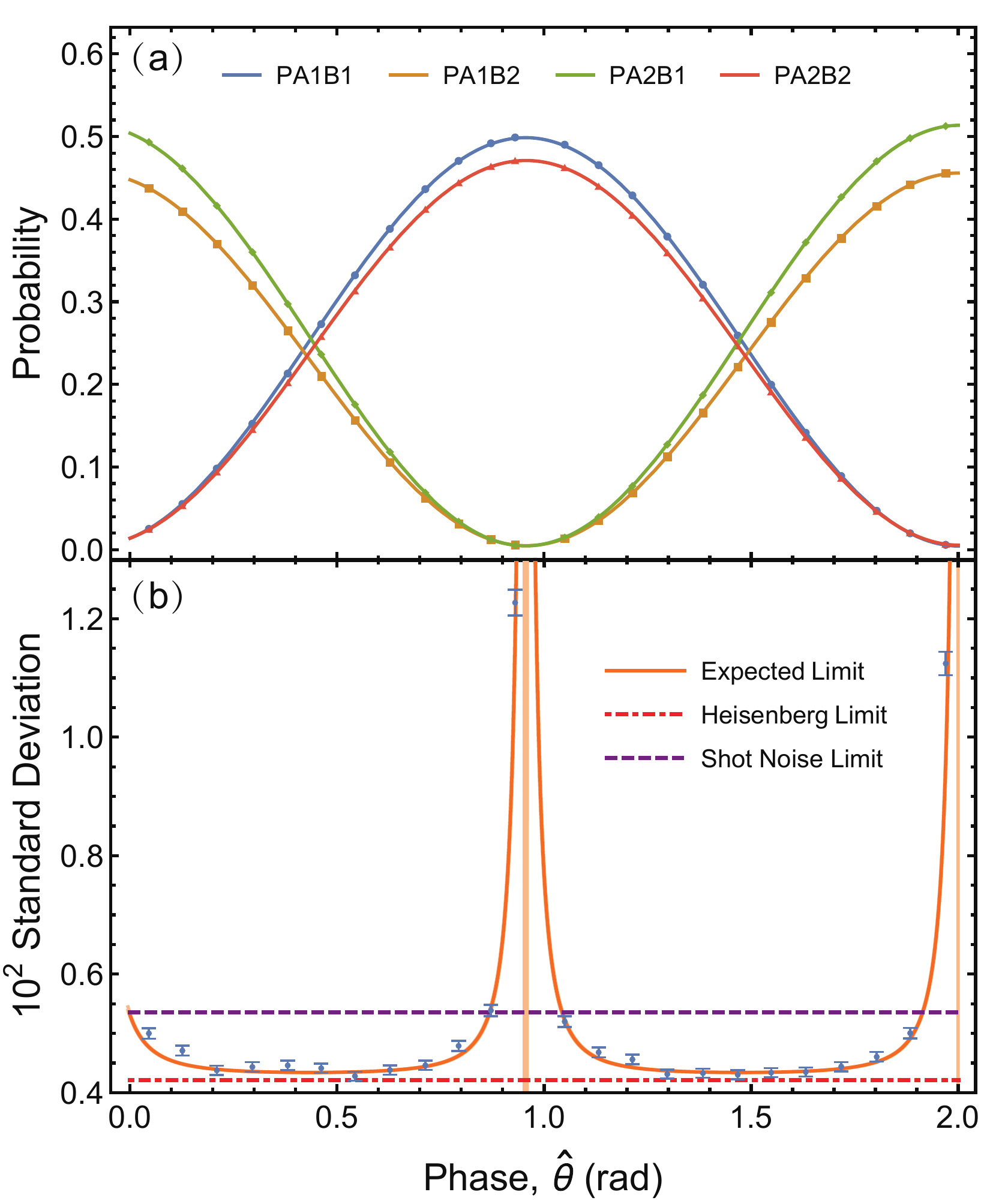}
	\caption{Experimental results for the distance of $240$ m.
\textbf{(a)} Experimental interference fringes of four detection events versus different phase shifts. The horizontal axis represents $\hat{\theta}=\frac{\theta_1-2\theta _2}{3}, \theta\in(0,\frac{2\pi}{3})$. \textbf{(b)} Experimental phase standard deviation (orange line) per trial versus different phase shifts. The shaded areas correspond to the $95\%$ confidence region, which are calculated from the uncertainty of the fit parameters. Purple dashed line: theoretical limit for SNL. Red dashed line: theoretical bound for HL. The error bars of the standard deviations are discussed in the main text.}
\label{fig:240m}
\end{figure}

\emph{Theory and experiments.}
We begin by introducing the basic tools and techniques used in this work. The scheme to estimate the linear function of independent phase shifts uses a combination of two techniques: entanglement and multiple sampling of the phase shift \cite{slussarenko2017unconditional,daryanoosh2018experimental,PhysRevLett.124.030501,higgins2007entanglement}. In a distributed multiphase quantum sensing network with $M$ nodes, the essential feature of multiple sampling of the phase shift is that the phase shift $\theta_{k}$ in each node $k$ is coherently accumulated over the $p_{k}$ uses of the local phase gate $U_{\theta_{k}}$ \cite{daryanoosh2018experimental,PhysRevLett.124.030501,higgins2007entanglement}. Thus, it promises to estimate any function(s) of $\{\theta_{k}\}$, e.g., $\sum_{k}p_k\theta_{k}$.

To demonstrate the distributed quantum sensing, we consider two separated quantum sensors Alice ($A$) and Bob ($B$) with two independent unknown phase shifts $\theta_{A}$ and $\theta_{B}$, and the global function $(normalized)$ $\hat{\theta}$ to be estimated is $\theta_{A}/3 - 2\theta_{B}/3$. The sensors $A$ and $B$ are entangled by the single state $1/\sqrt{2}(|HV\rangle - |VH\rangle)$, and  the local unitary evolution for each sensor is set to be $U_{\theta_{A}}=e^{-i\theta_{A}\sigma_{z}^{A}/2}$ and $U_{\theta_{B}}=e^{-i\theta_{B}\sigma_{z}^{B}/2}$ respectively. To estimate $\hat{\theta}$, the overall evolution $U_{\hat{\theta}}$ should be given by  $U_{\hat{\theta}}=U_{\theta_{A}}\otimes (U_{\theta_{B}}^2)$, where $U_{\theta_{B}}^2$ represents two applications of $U_{\theta_{B}}$ or that the photon in sensor B passes the phase gate twice. Therefore, the entanglement state after the evolution becomes $1/\sqrt{2}(|HV\rangle - e^{i3\hat{\theta}} |VH\rangle)$. Theoretically, by implementing the $\sigma_x$ basis measurement in sensors $A$ and $B$, the probabilities that can be observed are $P_{A_1B_1}=P_{A_2B_2}=(1-\cos(3\hat{\theta}))/4$ and $P_{A_1B_2}=P_{A_2B_1}=(1+\cos(3\hat{\theta}))/4$, where $A_iB_j$ represents two-photon coincidence events of detectors $A_i$ and $B_j$ for $i,j \in \{1,2\}$. To quantify the phase sensitivity, we use the  Cramer-Rao bound \cite{giovannetti2011advances} $\Delta \hat{\theta} \geq \frac{\bm{\alpha^T}\bm{\alpha}}{\sqrt{k \bm{\alpha^T}\bm{\text{F}}\bm{\alpha}}}$, where $k$ denotes the number of independent measurements, and $\alpha=(1/3,-2/3)$ is the coefficient vector of phase shift. $\bm{\text{F}}$ denotes the classical Fisher matrix with elements $(\bm{\text{F}})_{kl}=\sum\nolimits_{i}P_{i}\left[(\partial/\partial\theta_{k})P_{i} \right]\left[(\partial/\partial\theta_{l})P_{i} \right]$ for $i\in \{A_1B_1, A_1B_2, A_2B_1, A_2B_2\}$. The effective Fisher information (FI), $F(\hat{\theta})$, which is used for evaluating the estimation sensitivity, is given by,
\begin{equation} \label{eq:FI}
F(\hat{\theta})=\frac{ \bm{\alpha^T}\bm{\text{F}}\bm{\alpha}}{(\bm{\alpha^T}\bm{\alpha})^2}.
\end{equation}

However, in practice, the entanglement states are generated  probabilistically  at  random  times  by  the spontaneous  parametric downconversion (SPDC) source. Owing to the imperfect transmission and detection efficiency $\eta$, some photons do not lead to detection. Furthermore, owing to higher-order SPDC events (the occasional simultaneous emission of 4, 6,\dots photons),  the resources are equivalent to multiple (2, 3,\dots) trials. Therefore, we will obtain fifteen types of detection events including: a) one-channel-click events: only one of the four channels clicks, $A_1$, $A_2$, $B_1$, and $B_2$;  b) twofold coincidence events: Alice and Bob both have one of the two channels click, $A_iB_j(\hat{\theta})$, $i,j \in \{1,2\}$, or Alice (Bob) has two clicks on her (his) side, $A_1A_2$ and $B_1B_2$; c) threefold coincidence events: any three of the four channels click, $A_1A_2B_1(\hat{\theta})$, $A_1A_2B_2(\hat{\theta})$, $A_1B_1B_2(\hat{\theta})$, $A_2B_1B_2(\hat{\theta})$; d) fourfold coincidence events: both of four channels click, $A_1A_2B_1B_2(\hat{\theta})$. We count $k$ such events to complete the protocol, and each detection event represents a recorded trial.

Because one-channel-click events and two of the twofold coincidence events, $A_1A_2$ and $B_1B_2$, do not yield information about the global function $\hat{\theta}$, only the rest of nine detection events can be used to estimate $\hat{\theta}$. By ignoring the events of no channel click, the quantity of useful events equals to the sum of the other nine types of events, which is defined as $C_{sum}(\hat{\theta})$. Because the quantity of threefold and fourfold coincidence events is usually very low, we mainly consider the probabilities of two-photon coincidence events. The probabilities of these four types of double match as a function of phase shift are given by $P(A_iB_j(\hat{\theta}))={C_{A_iB_j}(\hat{\theta})}/{C_{sum}(\hat{\theta})}, i,j \in \{1,2\}$, where $C_{A_iB_j}(\hat{\theta})$ denotes the quantity of two-photon coincidence events. In the experiment, the projective measurements are performed in the $\sigma_x$ basis, which can achieve the maximum visibility for the interference fringe.

Fig.~\ref{fig:setup} (b)-(d) show the experimental diagram from entangled photon-pair emission to detection process. The two sensors are named Alice and Bob. The actual fiber distance between two sensors is $240$ m ($10$ km after adding the spools). The pump photons are injected into the periodically poled potassium titanyl phosphate (PPKTP) crystal in a Sagnac loop. For the pump lasers with the wavelength of $780$ nm, pulse width of $10$ ns, and frequency of $4$ MHz, the polarization-entangled photon pairs are generated from the loop at $1560$ nm. According to the theory, we create the maximally polarization-entangled two-photon state $\ket{\phi}=\frac{1}{\sqrt{2}}(\ket{HV}-\ket{VH})$. Then, the photon pairs are sent through fibers to two sensors for phase shifts and measurements. In Alice, one phase parameter is introduced, whereby injecting each photon into three plates, i.e., quarter-wave plate (QWP), half-wave plate (HWP), and quarter-wave plate, in sequence one time; the phase shift is denoted as $e^{i\theta_A}$. While in Bob, there is a loop, and each photon passes through the same plate group two times; thus the number of photon-passes \cite{higgins2007entanglement} is two in each trial, and the phase shift is $e^{i2\theta_B}$. We implement the entanglement state $\ket{\phi}=\frac{1}{\sqrt{2}}(\ket{HV}-e^{i3\hat{\theta}}\ket{VH})$ with a linear function defined as $\hat{\theta}=\frac{\theta_A-2\theta _B}{3}$; then, we perform the $\sigma_x$ basis measurement using the combination of HWP and polarizing beam splitter (PBS). After performing the detection procedure using superconducting nanowire single-photon detectors (SNSPDs), a time-digital convertor (TDC) is used to record the single photon detection events. Both Alice and Bob have two channels after PBS, including the reflected channel (ch1) and transmitted channel (ch2).

For the fair comparison with the SNL, we need to accurately count the quantity of resources \cite{slussarenko2017unconditional,resch2007time}. Owing to the imperfect transmission and detection efficiency, the actual number of photons (number of photons passing phase shifts) is larger than the number of recorded photons. Considering all the losses and imperfections, the actual number of photons $\widetilde{N}_{i}$ is related to the recorded number of photons ${N}_{i}$ by (see Appendix~\ref{Accurate statistics on the photon resources} for more details)
\begin{equation}
\widetilde{N}_{i}=\frac{{N}_{i}}{\eta_i}\times\frac{(4+\mu)\eta_i-4(2+\mu)}{2(2+\mu)(\eta_i-2)},
\end{equation}
where $i\in \{A_1, A_2, B_1, B_2\}$. Because the ideal classical scheme is assumed to be lossless and use all photon resources passing the phase shifts, it must be attributed to the effective number of resources. In our distributed multiphase quantum sensing, the effective number of resources is $n=\widetilde{N}_{A_1}+\widetilde{N}_{A_2}+2\widetilde{N}_{B_1}+2\widetilde{N}_{B_2}$. The SNL is achievable that Alice and Bob locally estimate their phase shift $\theta_{A}$ and $\theta_{B}$, respectively. Because the global function to be estimated is $\theta_{A}/3 - 2\theta_{B}/3$, the optimal strategy is to send $n/3$ photons to Alice, and Bob uses the rest ($2n/3$) photons to estimate $\theta_{B}$ without multipassing. Therefore, the SNL is obtained by $\sqrt{\frac{1}{9} \frac{3}{n}+ \frac{4}{9} \frac{3}{2n}}$, i.e., $1/ \sqrt{n}$. By establishing a theoretical model (see Appendix~\ref{Theoretical model}), we anticipate that a violation of SNL can be observed with an overall threshold efficiency of $\eta_{min}\approx 57.7\%$, which imposes rigorous limits on the sensing scales.

To realize the unconditional violation of SNL, we develop an entangled photon source with high heralding efficiency and high visibility. The beam waist of the pump beam ($780$ nm) is set to be $180$ $\upmu$m; then the beam waist of the created beam ($1560$ nm) is set to be $85$ $\upmu$m, which optimizes the efficiency of coupling $1560$ nm entangled photons into the single mode optical fiber, and the coupling efficiency is approximately $92.3\%$. The transmission efficiency for entangled photons passing through all optical elements in the source is approximately $95.9\%$. In the Sagnac loop, the clockwise and anticlockwise paths are highly overlapped, and we measure the visibility of the maximally polarization-entangled two-photon state to be $98.11\%$ with the mean photon number $\mu=0.0025$ to suppress the multi-photon effect. Using the superconducting nanowire single-photon detectors with high efficiencies of more than $92.2\%$, we achieved an average overall heralding efficiency of $73.88\%$. To perform the field test, we apply clock synchronization between the source and sensors with the repetition rate of $100$ kHz to guarantee the procedure of distributed quantum sensing. Because the outer environmental change leads to the irregular vibration of fibers, the stability of the field-test system is worse than that of the laboratory system. Thus, before collecting data, we need to calibrate the system to ensure that the photon polarization will not be affected by fibers. In addition, we overcome the problem of efficiency instability by tightly placing optical elements to reduce the light path (especially for the loop in Bob) and increasing the repetition rate to $4$ MHz to shorten the data collection time.

\emph{Results.}
The experiment is implemented with different fiber distances between sensors. First, we consider the distance of $240$ m. For each phase where $\hat{\theta}\in(0,\frac{2\pi}{3})$, there are approximately 9,500,000 recorded trials being collected to depict the interference fringes. The efficiencies of $A_1$, $A_2$, $B_1$, $B_2$ are $74.32$, $76.67$, $74.77$, and $69.74\%$, respectively. As shown in Fig.~\ref{fig:240m} (a), the interference fringe visibilities of $A_1B_1$, $A_1B_2$, $A_2B_1$, and
$A_2B_2$ are $98.27$, $97.93$, $98.23$, and $97.74\%$, respectively. After adding $10$ km spools between Alice and Bob, for each phase, where $\hat{\theta}\in(0,\frac{2\pi}{3})$, we collect approximately 7,000,000 recorded trials to portray the interference fringes. The efficiencies are $58.10$, $60.46$, $58.37$, and $52.84\%$, and the interference fringe visibilities are $96.50$, $94.86$, $95.60$, and $96.48\%$, respectively, as shown in Appendix~\ref{Interference fringes}. Then, we can obtain the Fisher information $F$ corresponding to each $\hat{\theta}$ from the interference fringes. The standard deviation of the estimate, $\delta \hat{\theta}$, is based on $\bar{k}$ measurement outcomes ($\bar{k}$ trials). To experimentally acquire the $\delta \hat{\theta}$, we repeat this $\bar{k}$ measurement $s$ times and obtain the distribution of $\hat{\theta}$.  For the distance of $240$ m, $s=1595$, and $\bar{k}$ is approximately $6200$ for each phase shift. For the distance of $10$ km, $s=1579$ and $\bar{k} \approx 4650$. The experimental Fisher information $F$ and $\delta \hat{\theta}$ has the relationship $\sqrt{F}=1/{(\delta \hat{\theta} \sqrt{\bar{k}})}$. The experimental error of $\delta \hat{\theta}$, $\Delta(\delta(\hat{\theta}))$, is well approximated by  $\Delta(\delta(\hat{\theta}))=\delta \hat{\theta}/\sqrt{2(s-1)}$ \cite{PhysRevLett.123.040501}, which is used for drawing the error bars. Considering the experimental device stability and substantial amount of data, the mean photon number $\mu$ is set to $0.056$ ($0.072$) for $240$ m ($10$ km), and the lowest efficiency of $240$ m exceeds the threshold efficiency. Our results demonstrate a wide range of violation of the SNL, and the system achieves the phase precision of $0.916$ dB below the corresponding SNL for the distance of $240$ m.

\begin{figure}
	\centering
	\includegraphics[width=0.45\textwidth]{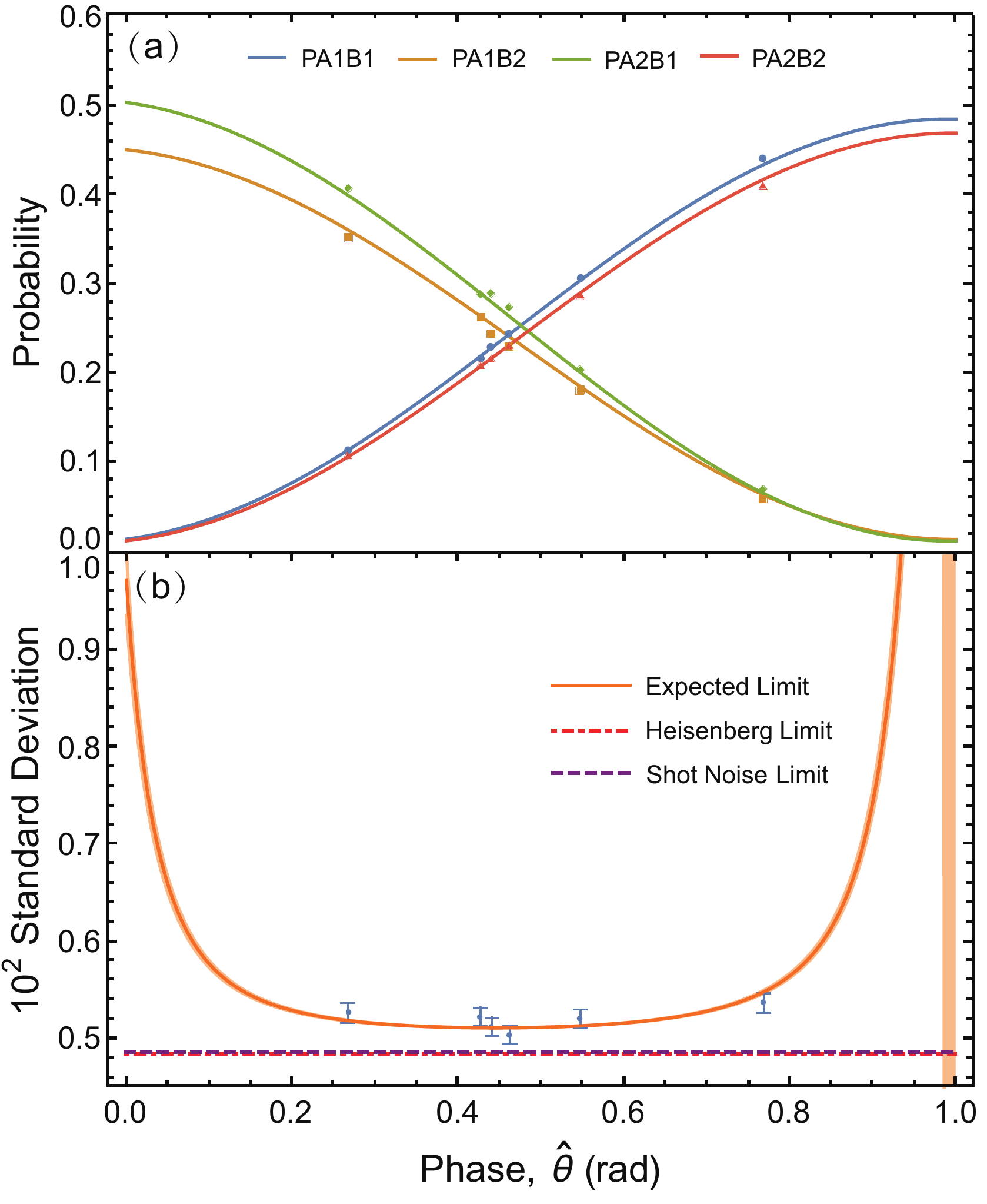}
	\caption{Experimentally measured phase estimate of the random detection events and phase standard deviations for the distance of $10$ km. \textbf{(a)} Experimental measurement of the random detection events, as shown by markers.  Lines represent the four probabilities from Appendix~\ref{Interference fringes}. \textbf{(b)} Experimental phase standard deviation per trial of random detection events. Orange line: expected phase standard deviation calculated from the Fisher information from Appendix~\ref{Interference fringes}. The shaded areas correspond to the $95\%$ confidence region, which are calculated from the uncertainty of the fit parameters. Purple dashed line: theoretical limit for SNL. Red dashed line: theoretical bound for HL. The error bars of the standard deviations are discussed in the main text.}
	\label{fig:random}
	
\end{figure}

One of the challenging problems in quantum metrology is to precisely measure the unknown phases. To simulate the actual situation of irregular phase transformation, the quantum random number generators (QRNGs) are employed to randomly control the rotation angle of HWP in Alice and Bob. A TDC is used to time-tag the random number signals of $200$ kHz, which is generated from local QRNG in real time. Then, the generated random bits `` $0$ '' and `` $1$ '' are divided into a group of $64$ bits; next, they are converted into decimal numbers, which control the rotating angles of HWP between $0$ and $2\pi$. We collect $6$ data points with different random phases, where $\hat{\theta}\in(0,\frac{\pi}{3})$, and accumulate approximately $7,500,000$ trials per phase. Fig.~\ref{fig:random} shows the experimentally measured probabilities of the random events and their calibration probability curves from Appendix~\ref{Interference fringes}. We repeat $\bar{k} \approx 4750$ measurement for $1579$ times, and acquire the standard deviations of the estimated phases, as listed in Tab.~\ref{tab:standard deviation}. The results can not beat the SNL due to the efficiency problem.
It is possible in the future that the lowest efficiency of $10$ km exceeds the threshold efficiency, so that the unknown phases could be precisely measured with the phase precision below the SNL.

\begin{table}[htbp]
\centering%
  \caption{Standard deviation of phase estimation. }
\begin{tabular}{c|c|c}
\hline
 & Estimated phase (rad) & Standard deviation ($\times 10^{-2}$)\\
\hline
1 & $0.768486$ & $0.536 \pm 0.010$ \\

2 & $0.547627$ & $0.520 \pm 0.009$ \\

3 & $0.462277$ & $0.503 \pm 0.009$ \\

4 & $0.440517$ & $0.511 \pm 0.009$ \\

5 & $0.267731$ & $0.526 \pm 0.009$ \\

6 & $0.427480$ & $0.522 \pm 0.009$ \\

\hline
\end{tabular}
\label{tab:standard deviation}
\end{table}

\emph{Conclusion.}
In summary, we demonstrated the realization of distributed multiphase quantum sensing with two remote sensors by considering the imperfections and losses of the system, taking all the photons used into account, and removing the post-selection of results. By utilizing the entanglement of quantum resource and appropriate measurements, the phase sensitivity, comparing with the SNL, has been distinctly enhanced. Furthermore, the unknown phases are measured with high accuracy. Our work advances the development of the quantum sensing network with more nodes and larger scale. We anticipate that this work will result in further improvements that will transition the quantum sensing network towards practical applications.

\begin{acknowledgements}
This work was supported by the National Key R\&D Program of China (Grants No. 2017YFA0303900, 2017YFA0304000), the National Natural Science Foundation of China, the Chinese Academy of Sciences (CAS), Shanghai Municipal Science and Technology Major Project (Grant No.2019SHZDZX01), and Anhui Initiative in
Quantum Information Technologies.

\end{acknowledgements}

\appendix

\section{Accurate statistics on the photon resources}\label{Accurate statistics on the photon resources}

Consider a SPDC source with a Poisson statistics,
\begin{equation}
P(n)=\frac{{\mu}^{n}}{n!} {e}^{-\mu},
\end{equation}
where $P(n)$ denotes the probability of generating $n$ pair of photons, and $\mu$ is the mean photon number. Due to the low mean photon number in the experiment, we estimate the resources with $1$ and $2$ pairs of photons, and ignore the higher-order (3, 4, 5,\dots) events. The ratio between $P(2)$ and $P(1)$ is $P(2)/P(1)=\mu/2$. The number of photons detected by each detector in Alice (Bob) is composed of three parts: a) one photon from $1$ pair of photons goes to the detector, and the probability is $\frac{1}{1+\mu/2}$; b) two photons from $2$ pair of photons go to the same detector, and the probability is $\frac{\mu/4}{1+\mu/2}$; c) two photons from $2$ pair of photons go to the different detector, and the probability is $\frac{\mu/4}{1+\mu/2}$. The actual number of photons $\widetilde{N}_{i}$ is related to the recorded number of photons ${N}_{i}$ by

\begin{align}
\widetilde{N}_{i}=&\frac{{N}_{i}}{\eta_i}\times\frac{1}{1+\mu/2}+\frac{{N}_{i}}{1-(1-\eta_i)^2}\times\frac{\mu/4}{1+\mu/2}\times2\nonumber\\
&+\frac{{N}_{i}}{\eta_i}\times\frac{\mu/4}{1+\mu/2}\nonumber\\
=&\frac{{N}_{i}}{\eta_i}\times\frac{(4+\mu)\eta_i-4(2+\mu)}{2(2+\mu)(\eta_i-2)},
\end{align}

where $i\in \{A_1, A_2, B_1, B_2\}$. The total number of  photon resources used in the experiment is $n=\widetilde{N}_{A_1}+\widetilde{N}_{A_2}+2\widetilde{N}_{B_1}+2\widetilde{N}_{B_2}$.

\section{Theoretical model}\label{Theoretical model}
In this section, we derive the overall threshold efficiency when a violation of SNL can be observed. The  single  state  after the evolution becomes $1/\sqrt{2}(|HV\rangle - e^{i3\hat{\theta}} |VH\rangle)$. Theoretically, by implementing the $\sigma_x$ basis measurement in Alice and Bob, the one-channel-click events $P_{A_1}$, $P_{A_2}$, $P_{B_1}$ and $P_{B_2}$ do not yield information about the global function. Therefore, we only need to consider the twofold coincidence events, and the probabilities that can be observed are $P_{A_1B_1}=P_{A_2B_2}={\eta}^2(1-\cos(3\hat{\theta}))/4$ and $P_{A_1B_2}=P_{A_2B_1}={\eta}^2(1+\cos(3\hat{\theta}))/4$, where $\eta$ is the efficiency of the detector. We  count $k$ such  events  to  complete  the  protocol,  and each detection event represents a recorded trial. By normalizing the four probabilities, the Fisher information of the global function $\hat{\theta}$ is $F(\hat{\theta})=9$. $N_{0}$ denotes total detection events, and we have $k=N_0{\eta}^2$. The standard deviation of $\hat{\theta}$ can be calculated as $\delta(\hat{\theta})=\sqrt{1/kF(\hat{\theta})}=1/(3\sqrt{N_{0}{\eta}^2})$. Since the total number of used photons is $3N_{0}$, the  SNL  is  obtained by $\delta_{SNL}(\hat{\theta})=1/\sqrt{3N_{0}}$. To beat the SNL, we should have ${\delta}(\hat{\theta})\le{\delta}_{SNL}(\hat{\theta})$, which requires $\eta\ge\frac{\sqrt{3}}{3}\approx0.577$.

\section{Interference fringes for the distance of $10$ km}\label{Interference fringes}

\begin{figure}[ht]
	\centering
	\includegraphics[width=0.42\textwidth]{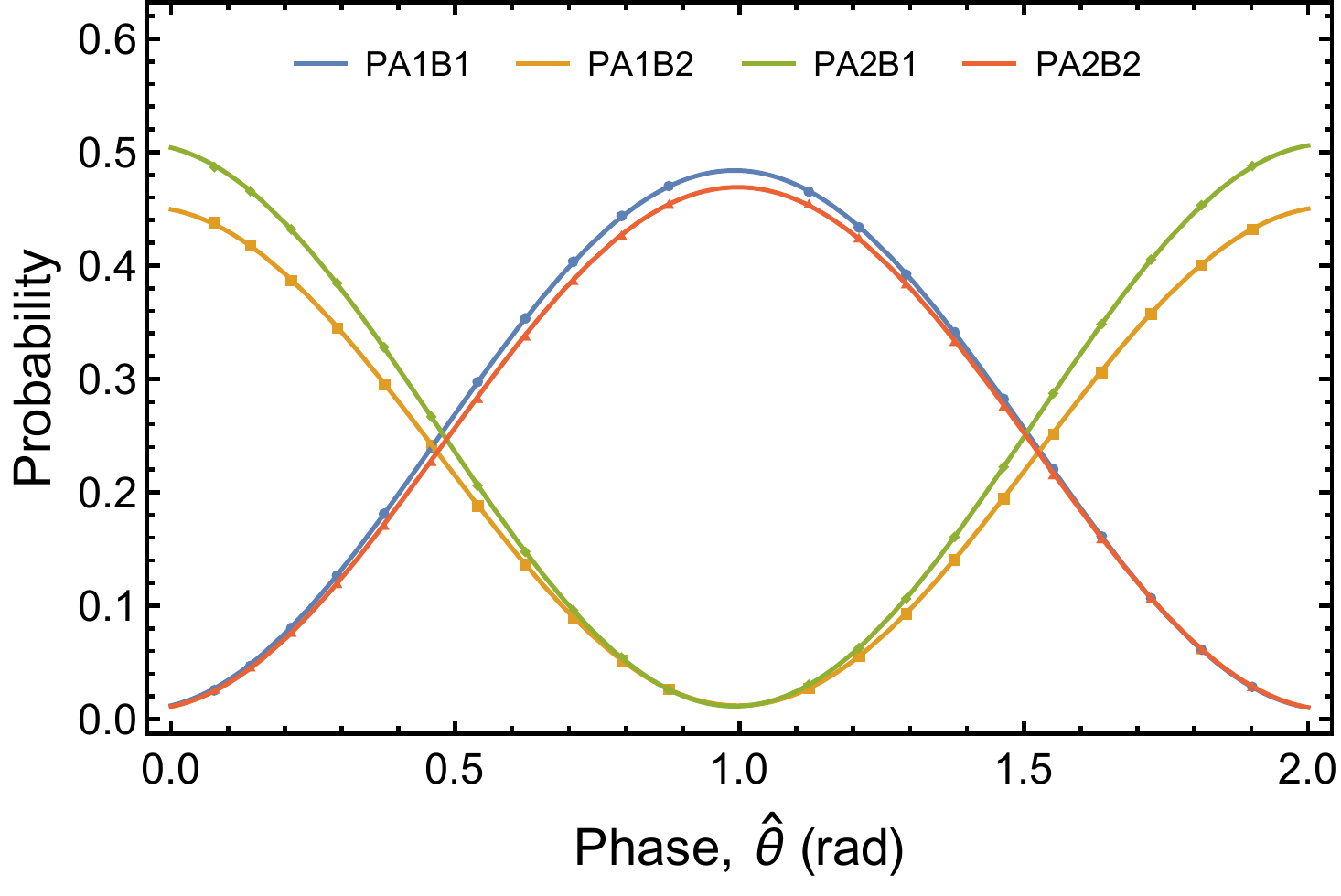}
	\caption{Experimental interference fringes of four detection events versus different phase shifts for the distance of $10$ km. The horizontal axis represents $\hat{\theta}=\frac{\theta_1-2\theta _2}{3}, \theta\in(0,\frac{2\pi}{3})$.}
\label{Fig:10km}
\end{figure}

The interference fringes for the distance of $10$ km are shown in Fig.~\ref{Fig:10km}.

\section{Determination of single photon efficiency}

The efficiencies are listed in the table below. The single photon heralding efficiency is defined as: $\eta_{A1}=C_{11}/N_{B1}$, $\eta_{B1}=C_{11}/N_{A1}$, $\eta_{A2}=C_{22}/N_{B2}$, $\eta_{B2}=C_{22}/N_{A2}$, where $C_{11}$ and $C_{22}$ denote the two-photon coincidence events about the reflected channels and transmitted channels, respectively. $N_{A1}$, $N_{B1}$ represent the single photon events of reflected channels, and $N_{A2}$, $N_{B2}$ represent the single photon events of transmitted channels.

The heralding efficiency is listed in Tab.~\ref{tab:efficiency}, where $\eta^{\mathrm{sc}}$ denotes the efficiency for entangled photons coupled into the single mode optical fiber,  $\eta^{\mathrm{so}}$ is the transmission efficiency for entangled photons crossing over optical elements in the source, $\eta^{\mathrm{fiber}}$ is the transmittance of a $120$ m fiber between the source and sensor, $\eta^{\mathrm{m}}$ is the efficiency for photons passing through the measurement apparatus, and $\eta^{\mathrm{det}}$ is the single photon detector efficiency.  The heralding efficiency and the transmittance of individual optical elements are listed in Tab.~\ref{tab:efficiency}, and the loss of $5$ km fiber spool in Alice (Bob) is $1.05$ dB ($1.14$ dB). Owing to the diameter limitation of HWP and QWP, there is an efficiency loss of approximately $4\%$ for light path clipping in Bob's loop.

\begin{center}
\begin{table}[htbp]
\footnotesize
\begin{tabular}{c|c|ccccc}
\hline
 & heralding efficiency ($\eta$) & $\eta^{\mathrm{sc}}$ & $\eta^{\mathrm{so}}$ & $\eta^{\mathrm{fiber}}$ & $\eta^{\mathrm{m}}$ & $\eta^{\mathrm{det}}$ \\
\hline
Alice1  & 74.32\% & 92.3\% & 95.9\% & 99.0\% & 91.0\% & 93.2\%\\
Bob1   & 74.77\% & 92.5\% & 95.9\% & 99.0\% & 87.5\% & 97.3\% \\
\hline
Alice2  & 76.67\% & 92.3\% & 95.9\% & 99.0\% & 91.6\% & 95.5\%\\
Bob2    & 69.74\% & 92.5\% & 95.9\% & 99.0\% & 86.1\% & 92.2\% \\
\hline
\end{tabular}
\caption{Characterization of optical efficiencies in the experiment.}\label{tab:efficiency}
\end{table}
\end{center}

\section{Tomography of quantum state}

\begin{figure}[htbp]
\centering
\resizebox{6cm}{!}{\includegraphics{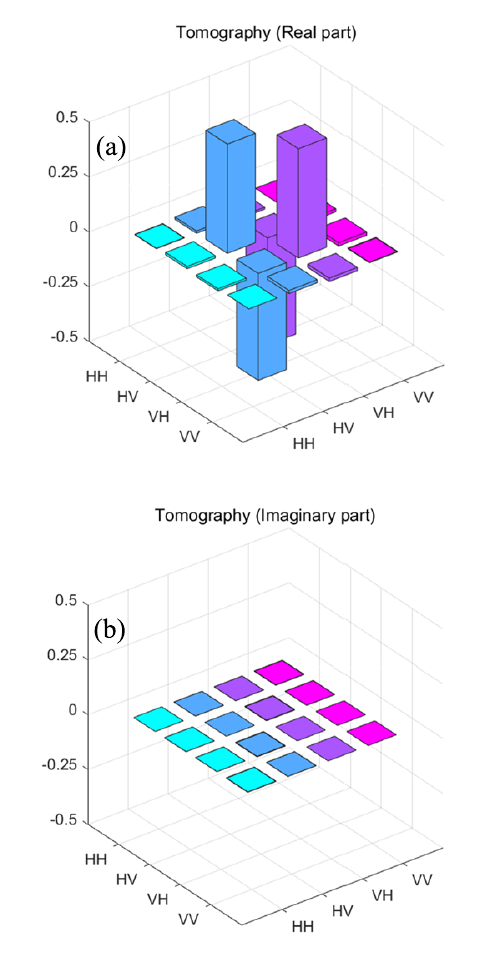}}
\caption{(color online) Tomography of the produced two-photon state. The real and imaginary components are demonstrated in (a) and (b), respectively.}
\label{Fig:tomo}
\end{figure}

In this experiment, we create the maximally polarization-entangled two-photon state $\ket{\phi}=\frac{1}{\sqrt{2}}(\ket{HV}-\ket{VH})$. The mean photon number is set as $\mu=0.0025$ to suppress the multi-photon effect in SPDC. The tomography measurement is performed of the entangled state with the fidelity of $98.58\%$, as shown in Fig.~\ref{Fig:tomo}. We assume that the imperfection originates from the multi-photon components, imperfect optical elements, and imperfect spatial/spectral mode matching.

\bibliography{SNL_V20}

\end{document}